# An Atom-Probe Tomographic Study of the Compositional Trajectories During γ(f.c.c.)- γ′(L1$_2$) Phase-Separation in a Ni-Al-Cr-Re Superalloy

Sung-Il Baik[1,2], Zugang Mao[1,2], Qingqiang Ren[1], Carelyn E. Campbell[3], Chuan Zhang[4], Bicheng Zhou[1,5], Ronald D. Noebe[6], David N. Seidman[1,2*]

[1]Department of Materials Science and Engineering, Northwestern University, Evanston, IL 60208, USA.

[2]Northwestern University Center for Atom Probe Tomography (NUCAPT), Evanston, IL 60208, USA

[3.] Materials Science and Engineering Division, National Institute of Standards and Technology (NIST), 100 Bureau Dr. Gaithersburg, MD 20899-8555, USA

[4.] CompuTherm LLC, 437 S. Yellowstone Dr., Madison, WI 53719, USA

[5.] Department of Materials Science and Engineering, University of Virginia, Charlottesville, VA, 22904, USA

[6.] NASA Glenn Research Center, 21000 Brookpark Rd, Cleveland, OH 44135, USA

*Corresponding author email address: *d-seidman@northwestern.edu*

The compositional diffusional trajectories for phase-separation of γ(face-centered-cubic (f.c.c.))-γ′(L1$_2$) phases are studied in a quaternary Ni–0.10Al-0.085Cr-0.02Re (mole fraction) alloy, aged at 700 ºC for 0 h to 1024 h, utilizing atom-probe tomography (APT). As the γ′(L1$_2$)-precipitates grow, Ni, Cr and Re become enriched and the Al is depleted near the interface on the γ(f.c.c.)-matrix-side as a result of diffusional fluxes crossing the γ(f.c.c)/γ′(L1$_2$) interface. The experimental (APT) compositional trajectories of the two-phases, γ(f.c.c.) and γ′(L1$_2$), are displayed in a quaternary phase-diagram, employing a tetrahedron, and compared with the Philippe- Voorhees (P-V) coarsening model, which includes the off-diagonal terms of the diffusion tensor for a multi-component system.

Nickel-based superalloys are widely used for commercial and military jet engines and land-based natural gas combustion turbines because of their superior mechanical properties, and creep and oxidation resistance at elevated temperatures [1-5]. The alloys are strengthened by coherent $\gamma'(L1_2)$-precipitates within a disordered Ni-rich $\gamma$(f.c.c.) matrix. During the last three decades, the mechanical properties of Ni-based superalloys operating at elevated temperatures have been remarkably improved, mainly due to the use of refractory elements [1, 6]. Rhenium, specifically, is of great interest because a small Re-addition improves their creep resistance considerably by increasing solid-solution strengthening of the $\gamma$(f.c.c.)-matrix and retarding the coarsening kinetics of $\gamma'(L1_2)$-precipitates at elevated temperatures [7-10].

In a dilute binary alloy, the coarsening of precipitates (Ostwald ripening) is described by the Lifshitz-Slyozov-Wagner (LSW) diffusion-controlled model [11, 12]. This model is, however, valid only for the near zero-volume faction of the second-phase because the diffusional interactions among precipitates are not taken into account. Kuehmann and Voorhees (K-V) [13] developed a model for Ostwald ripening of a ternary alloy by taking into account thermodynamic factors and kinetic effects. The development of atom-probe tomography (APT) [14-17] enables an understanding of coarsening behavior because it can measure the compositional trajectories and the temporal development of concentration profiles associated with $\gamma$(f.c.c.)/$\gamma'(L1_2)$ interfaces [18-21]. The K-V model doesn't, however, reproduce exactly the compositional trajectories of the $\gamma'(L1_2)$-precipitates because, it does not take into account the couplings among diffusional fluxes, which results from the minute details of the diffusion mechanism. [13]. This problem for multicomponent alloys is corrected by the Philippe-Voorhees (P-V) model [22], which includes the off-diagonal terms in the mobility tensor.

Herein, we present detailed results concerning $\gamma$(f.c.c.)-$\gamma'(L1_2)$ phase-separation and the temporal-dependent compositional trajectories for a quaternary Ni-Cr-Al-Re alloy utilizing APT. The effect of Re, a slow diffuser in Ni-Al-Cr alloys, on the temporal evolution of the compositional trajectories is represented utilizing a tetrahedron for a quaternary phase-diagram for the first time.

The Ni-based superalloy was prepared by induction-melting of relatively high purity elemental constituents under a partial Ar atmosphere and chill-cast in a copper mold to form a polycrystalline master ingot with a target composition of Ni-0.10Al-0.085Cr-0.022Re (mole fraction). The cast ingot was fully homogenized in the $\gamma$(f.c.c.)-phase field at 1300 °C for 20 h in vacuum and furnace cooled. The ingot was then sectioned into 1 cm thick slices, which were re-solutioned at 980 °C for 4 h in a drop-quench furnace and immediately water quenched without exposure to atmospheric pressure. The final solutionizing temperature is based on the differential thermal analysis (DTA) solvus temperature, ≈ 922 °C, which was performed on homogenized samples, at a rate of 10 K min$^{-1}$ in a helium atmosphere, cycled twice through the temperature range of the reaction. Finally, as-quenched samples were aged in the γ(f.c.c.) plus γ′(L1$_2$)-phase-field at 700 °C, for times ranging from 0.25 h to 1024 h in flowing argon, followed immediately by a quench into ice-brine water. APT nanotip specimens were cut from each of the aged samples and sharpened by a two-step electro-polishing procedure; 10% by volume perchloric acid in acetic acid and 2 vol.% perchloric acid in butoxyethanol at (5 to 21) V dc [23].

The 3-D APT experiments were performed utilizing a local-electrode atom-probe tomograph, LEAP5000XS[*], which has an 80 % detection efficiency [24]: Cameca Instruments, Inc., Madison, WI. The experiments were performed using voltage pulses at a pulse-fraction [(pulse-voltage)/(stationary DC voltage)] of 15 %, a pulse repetition rate of 250 kHz, a target detection rate of 0.02 ions pulse$^{-1}$, and a specimen temperature of (30.0 ± 0.2) K in to obtain accurate compositional measurements [25-27]. The recorded 3-D data were analyzed using the program IVAS 3.8.2 (Cameca Instruments).

**Figure 1** displays the temporal evolution of the γ′(L1$_2$)-precipitate morphology using 3-D APT reconstructions. The red 0.14 mole fraction Al iso-concentration surfaces [28] yield a visual comparison of the γ′(L1$_2$)-precipitates as a function of aging time. Very fine, Al-rich γ′(L1$_2$)-precipitates are observed even in the water-quenched sample after homogenization at 980 °C. This temperature is above the solvus temperature based on DTA measurements at 922 °C and a *Thermo-Calc* assessment at 915 °C using Chang's Ni database [8]. The presence of γ′(L1$_2$)-precipitates is attributed to the large solute supersaturations in the γ(f.c.c.)-matrix, which form during quenching from the homogenization temperature. Spheroidal γ′(L1$_2$)-precipitates appear in the as-quenched state, with a mean radius <R(t = 0 h)> = (1.06 ± 0.25) nm, which then grow and coarsen temporally to <R(t = 1024 h)> = (24.8 ± 6.63) nm, which is a factor of 23.4 increase. For the aging times investigated, the nanoscale γ′(L1$_2$)-precipitates remain spheroidal, indicating a small lattice parameter misfit between the two-phases with a concomitant small interfacial free energy between the γ(f.c.c.)/γ′L1$_2$)-phases [29-32]. The interfacial free energy in the current Ni-Al-Cr-Re alloy at

---

[*] Commercial names are used for completeness and do not constitute an endorsement from National Institute of Standards and Technology (NIST).

700 °C is estimated to be (18.6 ± 5.4) mJ/m$^2$ using the P-V model: see below for details. This value is significantly smaller than the value reported for our studies of binary Ni–Al alloys, (28.6 ± 1.6) mJ m$^{-1}$ [33, 34] and ternary Ni–Cr–Al alloys, (22 ± 7) mJ m$^{-1}$ [31, 35]. The number density, $N_v(t)$, of γ′(L1$_2$)-precipitates for the aging times investigated is a maximum at 0.25 h [$N_v(t)$ = (83.2 ± 7.90) x 10$^{22}$ m$^{-3}$] and it decreases continuously with increasing aging time [$N_v$ (t = 1024h) = (0.27 ± 0.06) x 10$^{22}$ m$^{-3}$]. The precipitate volume fraction, ϕ(t), measured by APT is asymptotically approaching a constant value after 16 h, (33.9 ± 3.6) %, which is the beginning of the quasi-stationary coarsening regime, where the volume fraction of the γ′(L1$_2$)-precipitates is approximately constant with aging time. The numerical values of <R(t)>, $N_v(t)$, and ϕ(t) of the γ′(L1$_2$)-precipitates, obtained from the 3-D APT experiments, are listed in **Table1**.

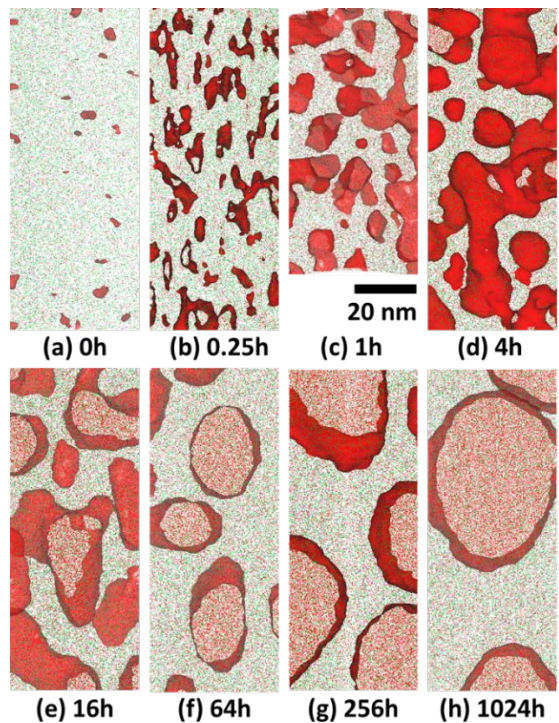

**Fig. 1.** Temporal evolution of the γ′(L1$_2$)-precipitate (red surface) in a quaternary Ni-0.10Al-0.085Cr-0.02Re alloy aged at 700 °C for times ranging from 0 h through 1024 h. Only a fraction (0.2%) of the Ni (green), Al (red), Cr (blue), and Re (dark yellow) atoms are displayed, for the sake of clarity, and the γ(f.c.c.)/γ′L1$_2$)-phases interfaces are delineated by red 0.14 mole fraction Al iso-concentration surfaces.

**Table 1.** The temporal evolution of the mean radius, $\langle R(t) \rangle$, number density, $N_v(t)$, volume fraction, $\phi(t)$, and their standard errors (two-sigma) for the γ'(L1$_2$)-precipitates in the Ni-0.1Al-0.085Cr-0.02Re (mole fraction) alloy, which was aged at 700 °C for times ranging from 0 h to 1024 h.

| Time (h) | $\langle R(t) \rangle$ (nm) | $N_v$ (10$^{22}$ m$^{-3}$) | $\phi(t)$ (%) |
|---|---|---|---|
| 0 | 1.06 ± 0.25 | 40.5 ± 7.48 | 0.38 ± 0.05 |
| 0.25 | 1.96 ± 0.34 | 83.2 ± 7.90 | 7.48% ± 0.71 |
| 1 | 3.15 ± 0.84 | 52.0 ± 4.72 | 18.39 ± 2.44 |
| 4 | 4.56 ± 1.73 | 18.0 ± 1.72 | 27.48 ± 2.68 |
| 16 | 7.13 ± 1.98 | 6.10 ± 0.83 | 33.94 ± 3.60 |
| 64 | 9.91 ± 3.07 | 2.24 ± 0.67 | 33.61 ± 5.21 |
| 256 | 14.6 ± 4.21 | 0.84 ± 0.15 | 32.35 ± 5.69 |
| 1024 | 24.8 ± 6.63 | 0.27 ± 0.06 | 32.74 ± 4.11 |

The compositions of the γ(f.c.c.)-matrix and the γ'(L1$_2$)-precipitate-phases evolve temporally and continuously as the γ'(L1$_2$)-precipitates become enriched in Al and depleted in Ni, Cr, and Re, **Fig. 2**. Each concentration profile across the γ(f.c.c.)/γ'L1$_2$)-interfaces was constructed employing a 0.2 nm bin size with respect to the 0.14 mole fraction Al iso-concentration surface utilizing the proximity histogram methodology [36], from which the standard error associated with the statistical variance (two-sigma) in the 3-D volumes was calculated [37]. The γ(f.c.c.)-matrix after homogenization has a composition of 0.782Ni-0.105Al-0.092Cr-0.021Re (mole fraction), while the γ'(L1$_2$)-nuclei have solute-supersaturated compositions of 0.727Ni-0.1945Al-0.0621Cr-0.0164Re (mole fraction), when <R(t = 0 h)> = (1.06 ± 0.25) nm. As the γ'(L1$_2$)-precipitates grow, the enrichments of Ni, Cr and Re, and depletions of Al in the γ(f.c.c.)-matrix develop as a result of diffusional fluxes crossing the γ(f.c.c.)/γ'(L1$_2$)-interface. The values the of solute excesses and depletions are represented by solute supersaturations ($\Delta \widetilde{C}_i$), which is the difference between the peak-value of the retention and the far-field composition (local equilibrium composition). The large initial Al-depletion in the γ(f.c.c.)-matrix reflects the fact that Al is the fastest-diffusing species, which is important for the nucleation and growth of γ'(L1$_2$)-precipitates. Rhenium, which has the smallest diffusivity in nickel, is supersaturated in both the γ(f.c.c.)- and γ'(L1$_2$)-phases, where it approaches approximately half of its equilibrium concentration when the Al concentration is close to its equilibrium value. The Ni supersaturation decreases slightly during the last aging stages as a result of the local balance among the fast Al-depletion and the late Cr and Re accumulations at the γ(f.c.c.)/γ′L1$_2$)-interface. After 4 h of aging time, the change in the solute supersaturations of all the elements is small ($d\Delta \widetilde{C}_i / dt \rightarrow 0$), implying the system is in a quasi-stationary coarsening regime slowly approaching its equilibrium volume fraction. The equilibrium γ(f.c.c.)-matrix and γ′(L1$_2$)-precipitate compositions at *infinite time* are obtained by extrapolating the measured concentrations from 4 h and are estimated to be 0.802Ni-0.051Al-0.1181Cr-0.0289Re (mole fraction) for the γ(f.c.c.)-matrix and 0.7584Ni-0.1763Al-0.0543Cr-0.011Re (mole fraction) for the γ′(L1$_2$)-precipitates.

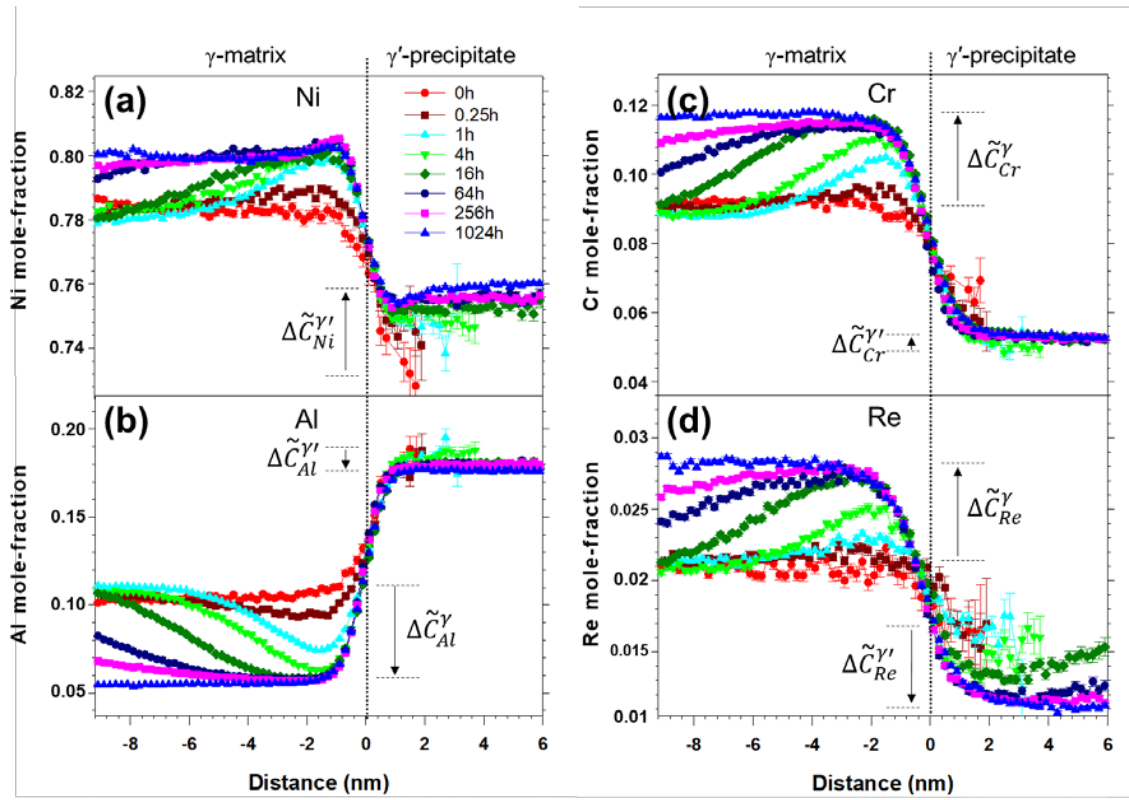

**Fig. 2.** The concentration profiles on either side of the γ(f.c.c.)/γ′(L1$_2$) interface of a quaternary Ni-Al-Cr-Re alloy aged at 973 K (700 ºC) for aging times from 0 h through 1024 h. The phase compositions evolve temporally, as the γ(f.c.c.)-matrix becomes enriched in Ni, Cr, and Re and depleted in Al. The dotted-vertical lines are placed at the inflection points of the Ni concentration-profiles indicating, by definition, the location of the γ(f.c.c.)/γ′(L1$_2$)-heterophase-interfaces. Note the indicated solute supersaturations ($\Delta \widetilde{C}_i$) of the elements Ni, Al, Cr, Re in the γ(f.c.c.)- and γ′(L1$_2$)-precipitate phases.

The compositional trajectories in local equilibrium across the γ(f.c.c.)/ γ′L1$_2$) interfaces are compared with a modified Gibbs-Thompson method approach developed by Philippe and Voorhees (P-V) [22]. They proposed a general model of coarsening for non-ideal and non-dilute solid-solutions, which includes off-diagonal terms in the diffusion tensor. The temporal evolution of the equilibrium composition in a multicomponent alloy is derived from the interfacial free energy, $\sigma$, determined from the following equation [22, 38]:

$$\sigma = \frac{9K(\Delta \bar{C}^{\gamma-\gamma'})^T M^{-1} \Delta \bar{C}^{\gamma-\gamma'}}{8V_m^{\gamma'}} \qquad (1)$$

where $K$ is the rate constant for the temporal evolution of <R(t)>, $\Delta \bar{C}^{\gamma-\gamma'}$ is the difference in the equilibrium compositions between the γ(f.c.c.)- and γ′(L1$_2$)-phases and the superscript letter T indicates the transpose of $\Delta \bar{C}^{\gamma-\gamma'}$, $M$ is the mobility tensor and $M^{-1}$ is its inverse, and $V_m^{\gamma'}$ is the molar volume of the γ′(L1$_2$)-phase, (6.7584 x 10$^{-6}$) nm$^3$ mol$^{-1}$ [10]. Based on the LSW and P-V coarsening models, the rate constant, $K$, obeys the relationship, $<R(t)>^3 = Kt$, where $K$ was determined using our experimental APT data, $(6.20 \pm 2.4 \times 10^{-30})$ m$^3$s$^{-1}$, for the current quaternary Ni-Al-Cr-Re alloy. The mobility tensor, $M$, is calculated from the relationship, $D = MG^\gamma$, where $D$ is the inter-diffusion tensor and $G^\gamma$ involves the second-derivatives of the Gibbs free energies in the γ(f.c.c.)-phase (a Hessian), which were calculated using TCNi8 and the Ni-mobility database [39] employing Thermo-Calc [40]:

$$G^\gamma = \begin{pmatrix} \dfrac{\partial G_\gamma^2}{\partial^2 C_{Al}} & \dfrac{\partial G_\gamma^2}{\partial C_{Al} \partial C_{Cr}} & \dfrac{\partial G_\gamma^2}{\partial C_{Al} \partial C_{Re}} \\ \dfrac{\partial G_\gamma^2}{\partial C_{Cr} \partial C_{Al}} & \dfrac{\partial G_\gamma^2}{\partial^2 C_{Cr}} & \dfrac{\partial G_\gamma^2}{\partial C_{Cr} \partial C_{Re}} \\ \dfrac{\partial G_\gamma^2}{\partial C_{Re} \partial C_{Al}} & \dfrac{\partial G_\gamma^2}{\partial C_{Re} \partial C_{Cr}} & \dfrac{\partial G_\gamma^2}{\partial^2 C_{Re}} \end{pmatrix} = \begin{pmatrix} 3.31 \times 10^5 & 1.34 \times 10^5 & 1.48 \times 10^5 \\ 1.20 \times 10^5 & 1.59 \times 10^5 & 0.12 \times 10^5 \\ 1.30 \times 10^5 & 0.08 \times 10^5 & 3.56 \times 10^5 \end{pmatrix} \text{ J·mol}^{-1} \quad (2)$$

and

$$D = \begin{pmatrix} D_{AlAl} & D_{AlCr} & D_{AlRe} \\ D_{CrAl} & D_{CrCr} & D_{CrRe} \\ D_{ReAl} & D_{ReCr} & D_{ReRe} \end{pmatrix} = \begin{pmatrix} 1.17 \times 10^{-18} & 4.68 \times 10^{-19} & 5.31 \times 10^{-19} \\ 1.35 \times 10^{-19} & 2.65 \times 10^{-19} & -9.38 \times 10^{-21} \\ -2.84 \times 10^{-20} & -1.54 \times 10^{-20} & -4.49 \times 10^{-21} \end{pmatrix} \text{ m}^2\text{s}^{-1} \qquad (3)$$

The reduction of the number of equations from $N$ to $N-1$ is obtained from the definition of the second-derivative of the Gibbs free energy as a dependent variable [22]. The positive values of the inter-diffusivities in $D$ indicate that the diffusion fluxes are causing phase separation, whereas the negative values indicate that the diffusional fluxes are moving from the precipitates to the disordered matrix. Employing the above values, the interfacial free energy, $\sigma$, is calculated using eq.(1) to be $(17.6 \pm 5.4)$ mJ/m$^2$ at 700 °C. Due to the addition of Re, $\sigma$ is about 21 % less than the value for Ni-Al-Cr alloys, as determined by APT experiments and first-principles calculations [41]. The value of $\sigma$ is also strongly influenced by the off-diagonal terms in the $G^\gamma$ and $D$ tensors, which are included in the P-V model. The compositional trajectory as a function of aging time in the γ(f.c.c.)-matrix is determined from [22],

$$C_i^\gamma(t) = C_i^{\gamma,eq} + (3\sigma V_m^{\gamma'})^{2/3} \Delta \overline{C}^{\gamma-\gamma'} \frac{[(\Delta \overline{C}^{\gamma-\gamma'})^T M^{-1} \Delta \overline{C}^{\gamma-\gamma'}]^{1/3}}{(\Delta \overline{C}^{\gamma-\gamma'})^T G^\gamma \Delta \overline{C}^{\gamma-\gamma'}} t^{-1/3} \qquad (4)$$

where $C_i^{\gamma,eq}$ are the far-field concentrations of elements in the γ(f.c.c)-matrix at infinite time. The γ(f.c.c.)-matrix supersaturations have the same temporal exponent, -1/3, in the LSW and P-V models, which have, however, a different amplitude for each element with a different rate constant. Because the amplitude is a product of a time-independent scalar and the difference of equilibrium compositions, the matrix supersaturations coincide with the tie-line representing the equilibrium conditions between the two-phases. The time-dependent composition of the γ′(L1$_2$)-phase precipitate is given by [22],

$$C_i^{\gamma'}(t) = C_i^{\gamma',eq} + (3\sigma V_m^{\gamma'})^{2/3} [(\Delta \overline{C}^{\gamma-\gamma'})^T M^{-1} \Delta \overline{C}^{\gamma-\gamma'}]^{1/3} \times \left\{ \frac{G^{\gamma'-1} G^\gamma \Delta \overline{C}^{\gamma-\gamma'}}{(\Delta \overline{C}^{\gamma-\gamma'})^T G^\gamma \Delta \overline{C}^{\gamma-\gamma'}} - \frac{G^{\gamma'-1} \Delta \overline{V}}{V_m^{\gamma'}} \right\} t^{-1/3} \qquad (5)$$

where $C_i^{\gamma',eq}$ is equilibrium concentrations of element, $i$, in the γ′(L1$_2$)-precipitates, $\Delta \overline{V}$ is the partial molar volume difference between the γ(f.c.c.)- and γ′(L1$_2$)-phases, $G^{\gamma'}$ is the second-derivative of the Gibbs free energies with respect to concentrations in the γ′(L1$_2$)-phase, which are obtained using TCNi8 employing Thermo-Calc [40]:

$$G^{\gamma'} = \begin{pmatrix} \dfrac{\partial G_{\gamma'}^2}{\partial^2 C_{Al}} & \dfrac{\partial G_{\gamma'}^2}{\partial C_{Al}\partial C_{Cr}} & \dfrac{\partial G_{\gamma'}^2}{\partial C_{Al}\partial C_{Re}} \\ \dfrac{\partial G_{\gamma'}^2}{\partial C_{Cr}\partial C_{Al}} & \dfrac{\partial G_{\gamma'}^2}{\partial^2 C_{Cr}} & \dfrac{\partial G_{\gamma'}^2}{\partial C_{Cr}\partial C_{Re}} \\ \dfrac{\partial G_{\gamma'}^2}{\partial C_{Re}\partial C_{Al}} & \dfrac{\partial G_{\gamma'}^2}{\partial C_{Re}\partial C_{Cr}} & \dfrac{\partial G_{\gamma'}^2}{\partial^2 C_{Re}} \end{pmatrix} = \begin{pmatrix} 3.05\times10^5 & 1.65\times10^5 & 2.07\times10^5 \\ 1.76\times10^5 & 4.82\times10^5 & 1.93\times10^5 \\ 2.09\times10^5 & 1.85\times10^5 & 10.2\times10^5 \end{pmatrix} \text{ J·mol}^{-1} \quad (6)$$

The calculated and measured compositions of the γ(f.c.c.)-matrix and γ′(L1$_2$)-precipitate phases are presented in a partial Ni-Al-Cr-Re quaternary phase-diagram utilizing a tetrahedron, **Fig. 3**. The blue- and red-colored surfaces represent the calculated conjugate solvus surfaces of the γ(f.c.c.)- and γ′(L1$_2$)-phases, respectively, utilizing *CompuTherm* [42]. The experimental data points (APT) indicated by the blue- and red-circles represent the γ(f.c.c.)- and γ′(L1$_2$)-phases' compositional trajectories, respectively, commencing with their initial compositions (t = 0) and ending at their final compositions (t = 1024). Solid-blue and red-circles are used to denote APT data in the quasi-stationary coarsening regime, t ≥ 16 h, and open-blue-circles and open-red-circles are used for t ≤ 4 h (Table 1). The compositions are calculated from the P-V model utilizing eqs. (4 and 5) and are represented by small solid-yellow circles. Because the P-V model is only for the quasi-stationary coarsening regime with an equilibrium volume fraction, the calculated points commence at an aging time of 16 h for both the γ(f.c.c.)- and γ′(L1$_2$)-phases, which are compared with the APT results employing blue- and red-circles, which start at 0 h (Table 1). In the P-V model, the compositional trajectories of the γ(f.c.c.)-matrix lie on a straight dashed tie-line (a vector) connecting the equilibrium concentrations of the γ(f.c.c.)- and γ′(L1$_2$)-phases at infinite time. Whereas, in the γ′(L1$_2$)-phase-field the composition trajectory of the γ′(L1$_2$)-precipitates does not lie on the equilibrium tie-line between the two phases, which is due to solute-flux coupling caused by the off-diagonal terms in the diffusion tensor.

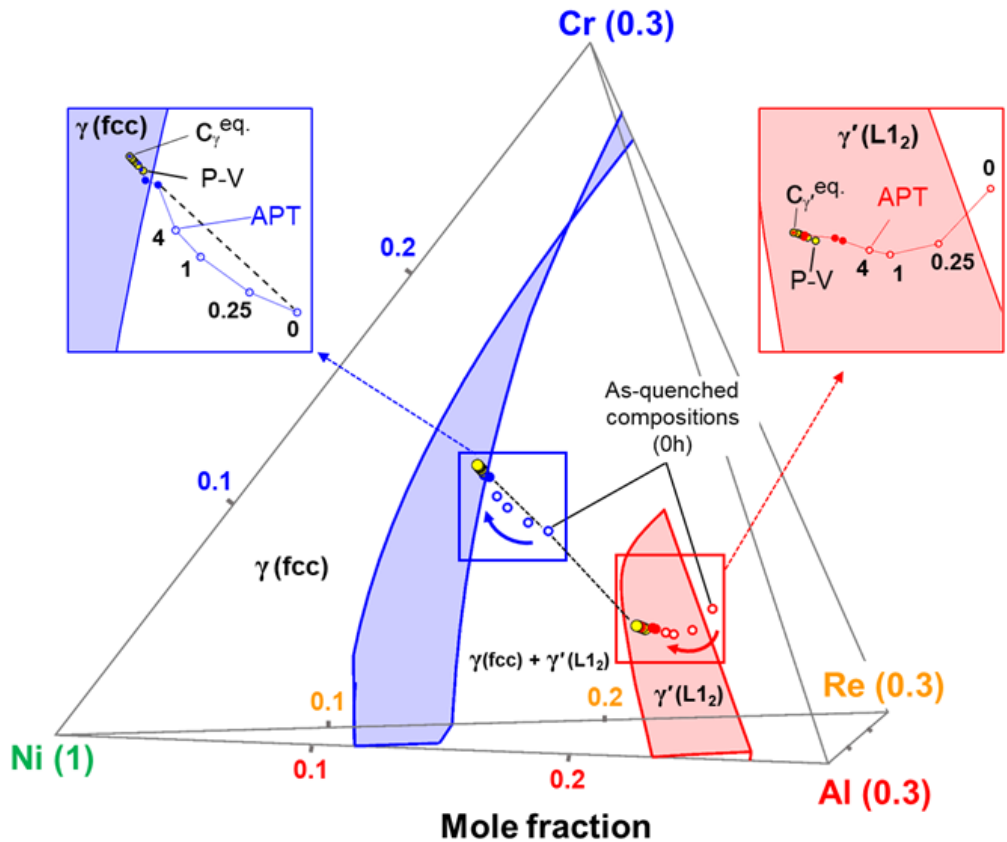

**Fig. 3.** A graphical representation of the composition trajectories for a quaternary Ni-Al-Cr-Re alloy, employing a tetrahedron, to display the experimental (APT) data and calculated values due to the P-V model. The equilibrium tie-line, at infinite time, connecting the conjugate solvus surfaces is indicated by a black dashed-line, which is a vector. All the experimental data points (APT) are indicated by the blue- and red-circles for the $\gamma$(f.c.c.)- and $\gamma'(L1_2)$-phases, respectively, which commence with their initial compositions (t = 0) and end at their final equilibrium compositions. The solid-blue- and solid-red-circles are for APT data in the quasi-stationary coarsening regime, t ≥ 16 h. And the open-blue- and open-red-circles are for t ≤ 4 h. The compositions calculated from the P-V model commencing at an aging time of 16 h are represented by solid-yellow circles. The behavior of the supersaturations in the $\gamma$(f.c.c.)-matrix and $\gamma'(L1_2)$-precipitates are magnified on the left- and right-hand-sides of the tetrahedron, respectively.

The different behaviors of the supersaturations in the γ(f.c.c.)-matrix and γ′(L1$_2$)-precipitates are presented in high-magnification images on the left- and right-hand-sides of the tetrahedron, respectively. Between 0 h and 16 h the volume fractions haven't yet achieved their quasi-stationary values, Table 1, because the nucleation and growth regimes are dominant. At times ≤4 h the APT experimental data points (represented by open-circles) lie on a curve in both phases, and not on a vector, because Al diffuses significantly faster than Cr and Re during the initial stages of phase-separation. The fastest diffusion behavior is well represented by the largest initial Al-depletion, $\Delta \widetilde{C}_{Al}^{\gamma}$, in the γ(f.c.c.)-matrix's concentration profile, Fig. 2. Whereas Re, which has the smallest diffusivity, approaches approximately half of its equilibrium concentration, $\Delta \widetilde{C}_{Re}^{\gamma}$, when the Al concentration is close to its equilibrium value. For times greater than 16 h, in the quasi-stationary coarsening regime, the supersaturations, $\Delta \widetilde{C}_i$, are very small in both phases and the concentration trajectories are close to a straight line (a vector), which agree with the quasi-stationary diffusional behaviors described by the P-V model [22]. This implies that the compositions of the γ′(L1$_2$)-precipitates in a quaternary system pertain to the thermodynamically equilibrium compositions including interfacial curvature (capillary effect [10-13, 22])). Nucleation and growth kinetics make, however, the compositional trajectories more complex, possibly leading to compositional variations within the precipitates; for example, in the Al-Sc-Zr system nanoprecipitates have a metastable core-shell structure [43-44].

In summary, the compositional diffusional trajectories in a phase-separating quaternary Ni–0.1Al-0.85Cr-0.02Re (mole fraction) alloy aged at 700 ºC for 0 h to 1024 h are studied utilizing the temporal evolutions of the concentration profiles at the γ(f.c.c.)/γ′(L1$_2$)-interface, utilizing proximity histograms, as determined by atom-probe tomography (APT). The local equilibrium concentrations across the γ(f.c.c.)/γ′(L1$_2$) interfaces are represented in a quaternary phase-diagram, employing a tetrahedron, and compared with the Philippe-Voorhees (P-V) coarsening model in the quasi-stationary coarsening regime. Initially, the small diffusivities of Cr and Re compared to Al's large diffusivity, results in a curvilinear behavior during the nucleation and growth regimes of the precipitates, which deviate from the equilibrium γ(f.c.c.)-γ′(L1$_2$) tie-line. In the quasi-stationary coarsening regime, >16 h, the concentration trajectories are close to a straight line (a vector), which follow the stationary diffusional behavior described by the P-V model for both the γ(f.c.c.) and

γ′(L1$_2$) phases. The calculated interfacial free energy, $\sigma$, is (17.6 ± 5.4) mJ/m$^2$ at 700 °C, for the γ(f.c.c.)/γ′(L1$_2$) interfaces, employing the mobility tensor in the P-V model.

## Acknowledgements

This research was supported by the National Science Foundation, Division of Materials Research (DMR) grant number DMR-1610367 001, Profs. Diana Farkas and Gary Shiflet, grant officers. Atom-probe tomography was performed at the Northwestern University Center for Atom-Probe Tomography (NUCAPT). The LEAP tomograph at NUCAPT was purchased and upgraded with grants from the NSF-MRI (DMR-0420532) and ONR-DURIP (N00014-0400798, N00014-0610539, N00014-0910781, N00014-1712870) programs. NUCAPT received support from the MRSEC program (NSF DMR-1720139) of the Materials Research Center, the SHyNE Resource (NSF ECCS-1542205), and the Initiative for Sustainability and Energy (ISEN) at Northwestern University. We also thank research associate professor Dr. Dieter Isheim for managing NUCAPT.